\documentclass[twocolumn,showpacs,amssymb,amsmath,aps,pra]{revtex4-1}
\newcommand{\half}{\mbox{$\frac{1}{2}$}}
\usepackage{bm}
\usepackage{graphics}
\usepackage{amsmath}
\usepackage{amssymb}

\begin{document}
\title{Even-odd entanglement in boson and spin systems}
\author{R. Rossignoli, N. Canosa, J.M. Matera}
\affiliation{Departamento de F\'{\i}sica-IFLP,
 Universidad Nacional de La Plata, C.C. 67, La Plata (1900), Argentina}
\date{\today}

\begin{abstract}
We examine the entanglement entropy of the even half of a translationally
invariant finite chain or lattice in its ground state. This entropy measures
the entanglement between the even and odd halves (each forming a ``comb'' of
$n/2$ sites) and can be expected to be extensive for short range couplings away
from criticality. We first consider bosonic systems with quadratic couplings,
where analytic expressions for arbitrary dimensions can be provided. The
bosonic treatment is then applied to finite spin chains and arrays by means of
the random phase approximation. Results for first neighbor anisotropic $XY$
couplings indicate that while at strong magnetic fields this entropy is strictly
extensive, at weak fields important deviations arise, stemming from
parity-breaking effects and the presence of a factorizing field (in which
vicinity it becomes size-independent and  identical to the entropy of a
contiguous half). Exact numerical results for small spin $s$ chains are shown
to be in agreement with the bosonic RPA prediction.
\end{abstract}
\maketitle
\section{Introduction}

The entanglement properties of many-body systems are of great interest for both
quantum information theory \cite{NC.00} and condensed matter physics
\cite{ON.02,AFOV.08,ECP.10}. Their knowledge enables, on the one hand, to
assess the potential of a given many-body system for quantum information
processing tasks such as quantum teleportation \cite{BE.93} and quantum
computation \cite{NC.00,JLV.03,RB.01}. On the other hand, it provides a deep
understanding of quantum correlations and their relation with criticality
\cite{ON.02,AFOV.08,ECP.10,OAFF.02,VLRK.03}. In non-critical systems with short
range couplings, i.e., local couplings in boson or spin lattices, ground state
entanglement is believed to satisfy a general area law by which the entropy of
the reduced state of a given region, which measures its entanglement with the
rest of the system, scales as the area of its boundary as the system size
increases \cite{ECP.10,PEDC.05}. This behavior is quite different from that of
standard thermodynamic entropy which scales as the volume. In one dimensional
systems this statement has been quite generally and rigorously proved
\cite{AEPW.02,ECP.10} and simply means that the entropy of a contiguous section
saturates, i.e., approaches a size independent constant, as the size increases.
Violation of this scaling is therefore an indication of criticality
\cite{VLRK.03,OAFF.02,CC.04}. The exact expression of the entropy of a
contiguous block in a one-dimensional XY spin 1/2 chain in the thermodynamic
limit has been obtained \cite{FIJK.07,IJK.05,PE.04} and confirms the previous
behavior.

The conventional area law holds for contiguous subsystems. For non-contiguous
regions it actually implies that the entropy is proportional to the number of
couplings broken by the partition. For instance, for {\it comb}-like regions
like the subset of all even sites in a chain, the entropy should scale as the
total number $n$ of sites for first neighbor or short range couplings. This was
in fact verified in \cite{AEPW.02} for the harmonic cyclic chain, where the
corresponding logarithmic negativity was calculated, and also verified
numerically in \cite{CWZ.06} for some spin arrays and a $1$-$d$ half-filled
Hubbard model, where the even entanglement entropy was computed. An exact
treatment of general comb entropies for a large one-dimensional critical XX
spin $1/2$ chain with first neighbor couplings was given in \cite{KMN.06},
showing that they are indeed proportional to the size $L$ plus a logarithmic
correction.

The aim of this work is to analyze in detail the entanglement entropy of all
even sites in {\it finite} boson and spin arrays, both in one dimension as well
as in general $d$-dimensions. Such bipartition can be normally expected to be
the maximally entangled bipartition at least for uniform nearest
neighbor couplings, as it will there break all coupling links. We first
analyze the bosonic case with general quadratic couplings, where a fully
analytic treatment of this entropy is shown to be feasible and allows to derive
simple general expressions in the weak coupling limit. Comparison with single
site and block entropies is also made. The bosonic treatment is then applied to
finite spin $s$ arrays with anisotropic ferromagnetic-type $XY$ couplings in a
uniform transverse field through the RPA approach \cite{MRC.10}. This allows to
predict in a simple way the main properties of the total even entropy in these
systems. Comparison with exact numerical results indicate that the RPA
prediction, while qualitatively correct, is also quite accurate outside the
critical region already for low spin $s\agt 2$, representing the high spin
limit. Results corroborate that for strong fields, the total even entropy in
these systems is extensive, i.e., directly proportional to the total number $n$
of sites. However, for low fields $B<B_c$, this entropy has an additive
constant, which arises in the RPA from parity restoration \cite{MRC.10}.
Moreover, in the immediate vicinity of the factorizing field $B_s<B_c$
\cite{KTM.82,AA.06,RCM.08,GAI.08}, extensivity is fully lost and the total even
entropy reduces to this constant, which is the same as that for the block
entropy and is exactly evaluated. The exact bosonic treatment is described in
sec.\ \ref{II}, whereas its application to spin systems is discussed in sec.\
\ref{III}. Conclusions are finally drawn in \ref{IV}.

\section{Entanglement entropy in bosonic systems\label{II}}
We start by considering a system of $n$ bosonic modes defined by boson creation
operators $b^\dagger_i$ ($[b_i,b^\dagger_j]=\delta_{ij}$), interacting through
a general quadratic coupling. The Hamiltonian can be written as
\begin{eqnarray}
H&=&\sum_{i,j}(\lambda_i\delta_{ij}-\Delta^+_{ij}) (b^\dagger_i b_j+\half
\delta_{ij})
 -\half (\Delta^-_{ij}b^\dagger_ib^\dagger_j+\bar{\Delta}^-_{ij}b_j b_i)
\label{H1}\\
&=&\half {\cal Z}^\dagger {\cal H} {\cal Z}\,,
 \;{\cal Z}=\left(\begin{array}{c}b\\b^\dagger\end{array}\right),
 \;{\cal H}=\left(\begin{array}{cc}\Lambda-\Delta^+&-\Delta^-\\
 -\bar{\Delta}^-&\Lambda-\bar{\Delta}^+\end{array}\right)\,,
 \nonumber\end{eqnarray}
where ${\cal Z}^\dagger=(b^\dagger,b)$, $\Lambda_{ij}=\lambda_i\delta_{ij}$ and
the $2n\times 2n$ matrix ${\cal H}$ is hermitian. The system is assumed stable,
such that the matrix ${\cal H}$ is {\it positive definite}. We may then also
write (\ref{H1}) in the standard diagonal form
\begin{equation}
H=\sum_{k}\omega_k ({b'}^\dagger_kb'_k+\half)\,,\label{Hd}
\end{equation}
where $\omega_k$ are the symplectic eigenvalues of ${\cal H}$, i.e., the
positive  eigenvalues of the matrix ${\cal M}{\cal H}$, with ${\cal
M}=(^{1\;\;0}_{0-1})$, which come in pairs of opposite sign and are all real
non-zero when ${\cal H}$ is positive definite \cite{RS.80}, and
${b'}^\dagger_k$ are the normal boson operators determined by the diagonalizing
Bogoliubov transformation \cite{RS.80} ${\cal Z}={\cal W}{\cal Z}'$ satisfying
${\cal W}^\dagger {\cal M}{\cal W}={\cal M}$ and $({\cal W}^\dagger {\cal
H}{\cal W})_{kk'}=\omega_k\delta_{kk'}$. The ground state is the vacuum
$|0'\rangle$ of the operators $b'_k$ and is non-degenerate.

Ground state entanglement properties can be evaluated through the general
Gaussian state formalism \cite{AEPW.02,CEPD.06,ASI.04}, which we here recast in
terms of the contraction matrix \cite{MRC.10,RS.80}
\begin{eqnarray}
{\cal D}&=&\langle {\cal Z}{\cal Z}^\dagger\rangle_{0'}-{\cal M}=
{\cal W}\left(\begin{array}{cc}0&0\\0&1\end{array}\right){\cal W}^\dagger\\
&=&\left(\begin{array}{cc}F^+&F^-\\
\bar{F}^-&I+\bar{F}^+\end{array}\right)\label{D},
\;\begin{array}{c}F^+_{ij}=\langle b^\dagger_j b_i\rangle_{0'}\\
F^-_{ij}=\langle b_j b_i\rangle_{0'}=
\langle b^\dagger_ib^\dagger_j\rangle_{0'}^*\end{array}\,.
\end{eqnarray}
This hermitian matrix determines, through application of Wick's theorem
\cite{RS.80}, the average of any many-body operator. In particular, the reduced
state $\rho_{A}={\rm Tr}_{\bar{A}}|0'\rangle\langle 0'|$ of a subsystem $A$ of
$n_A$ modes ($\bar{A}$ denoting the complementary subsystem and ${\rm
Tr}_{\bar{A}}$ the partial trace) is fully determined by the corresponding
sub-matrix ${\cal D}_A =\langle Z_AZ_A^\dagger\rangle-{\cal M}_A$ (Eq.\
(\ref{D}) with $i,j\in A$) and can be written as \cite{MRC.10}
\begin{equation}
\rho_A=\exp[-\half Z^\dagger_A\tilde{\cal H}_A Z_A] /{\rm Tr}\exp[-\half
Z^\dagger_A{\cal H}_A Z_A]\,,
 \label{rhoa}\end{equation}
where $\tilde{\cal H}_A={\cal M}_A\ln[I+{\cal M}_A{\cal D}_A^{-1}]$. Eq.\
(\ref{rhoa}) represents a thermal-like state of suitable $n_A$ independent
modes determined by the effective Hamiltonian $\tilde{\cal H}_A$. The
entanglement entropy of the $(A,\bar{A})$ partition,
$S(\rho_A)=S(\rho_{\bar{A}})$, is then determined by the symplectic eigenvalues
$f^A_k$ of ${\cal D}_A$ (i.e., the positive eigenvalues of the matrix ${\cal
D}_A{\cal M}_A$, which has eigenvalues $f_k^A$ and $-1-f_k^A$), and given by
\begin{eqnarray}
S(\rho_A)&=&-{\rm Tr}\,\rho_A\ln\rho_A=\sum_{k=1}^{n_A} h(f_k^A)\,,\label{SA}\\
h(f)&=&-f\ln f+(1+f)\ln(1+f)\label{Hf}\,.
 \end{eqnarray}

For instance, the entanglement of a {\it single mode} $i$ with the rest of the
system is just
\begin{eqnarray}
S(\rho_i)&=&h(f_i)\,,\label{Si}\;
f_i=\sqrt{(F^+_{ii}+\half)^2-|F^-_{ii}|^2}-\half\label{fi}\,,
\end{eqnarray}
where $f_i$, the symplectic eigenvalue of the single mode contraction matrix
${\cal D}_i$, represents the deviation from minimum uncertainty of the mode:
$(F^+_{ii}+\half)^2-|F^-_{ii}|^2=\langle q_i^2\rangle_{0'}\langle
p_i^2\rangle_{0'} -[{\rm Re}(\langle q_i p_i\rangle_{0'})]^2\geq 0$ for
$q_i=\frac{b_i+b^\dagger_i}{\sqrt{2}}$,
$p_i=\frac{b_i-b^\dagger_i}{\sqrt{2}i}$.

\subsection{Finite translationally invariant systems}
Let us now associate each bosonic mode with a given site in a cyclic chain and
consider a translationally invariant system of $n$ sites, such that
$\lambda_i=\lambda$ and $\Delta^{\pm}_{ij}=\Delta^{\pm}(i-j)$, with
$\Delta^{\pm}(-l)=\Delta^{\pm}(n-l)$. We first consider for simplicity the
one-dimensional case. Through a discrete Fourier transform
$b^\dagger_i=\frac{1}{\sqrt{n}}\sum_{k=0}^{n-1}e^{i2\pi ki/n}{b}^\dagger_k$, we
can diagonalize ${\cal H}$ analytically and obtain an explicit expression for
the contractions $F^{\pm}_{ij}$. We will assume
$\Delta^{\pm}(l)=\Delta^{\pm}(-l)$ $\forall$ $l$, in which case the energies
$\omega_k$ in (\ref{Hd}) adopt the simple form \cite{MRC.10}
\begin{eqnarray}\omega_k&=&\sqrt{(\lambda-\Delta^+_k)^2-(\Delta^-_k)^2}
 \label{wk}\,,\end{eqnarray}
where $\Delta^\pm_k$ are the Fourier transforms of the couplings:
\begin{eqnarray}
\Delta^{\pm}_k&=&\sum_{l=0}^{n-1}e^{i2\pi kl/n}\Delta^{\pm}(l)\,.\label{Dk}
\end{eqnarray}
The contractions $F^\pm_{ij}$ depend just on the separation $l\equiv |i-j|$ and
are given by
\begin{eqnarray}
F^{\pm}_{l}&\equiv&F^{\pm}_{j+l,j}
=\frac{1}{n}\sum_{k=0}^{n-1} e^{-i 2\pi kl/n} f^{\pm}_k\,,
\label{Fk}\\
f^{+}_k&=&\langle b^\dagger_k b_k\rangle_{0'}=\frac{\lambda-\Delta^+_k}
{2\omega_k}-\half,\; f^-_k=\langle b_k
b_{-k}\rangle_{0'}=\frac{\Delta^-_k}{2\omega_k}\,.\label{Fpm}
\end{eqnarray}
The symplectic eigenvalues of the full contraction matrix (\ref{D}) are of
course $f_k=\sqrt{(\half+f^+_k)^2-(f^-_k)^2}-\half=0$ $\forall$ $k$.

In the {\it weak coupling limit} $|\Delta^{\pm}_k|\ll \lambda$ $\forall$ $k$,
$f_k^\pm$ become small and up to lowest non-zero order we obtain
\begin{equation}
f^-_k\approx \frac{\Delta^-_k}{2\lambda},\;\;
f^+_k\approx\frac{(\Delta^-_k)^2}{4\lambda^2}\approx(f_k^-)^2\,,
 \label{wekk}\end{equation}
which leads to
\begin{eqnarray} F^{-}_l\approx \frac{\Delta^-(l)}{2\lambda},\;
F^+_l\approx\frac{\sum_{l'}\Delta^-(l')\Delta^-(l-l')}{4\lambda^2}\,.
\label{wek}
\end{eqnarray}
At this order just sites linked by $\Delta^-(l)$ or its convolution are
correlated. The eigenvalues $f_k^A$ of subsystem contraction matrices will
depend up to lowest non-zero order on $F^+_l$ and $(F^-_l)^2$, being then
$O(\Delta_-^2/\lambda^2)$ for $\Delta^-(l)\propto \Delta_-$.  We can then use
in (\ref{SA}) the approximation
\begin{equation}h(f)\approx -f(\ln f -1)+O(f^2)\,,\label{Hw}\end{equation}
such that $S(\rho_A)=O(\frac{\Delta_-^2}{\lambda^2}\ln
\frac{\Delta_-^2}{\lambda^2})$.

On the other hand, it is seen from Eq.\ (\ref{wk}) that the present system is
stable provided $\lambda\geq \Delta^+_k+|\Delta^-_k|$ $\forall$ $k$. For {\it
attractive} couplings $\Delta^{+}(l)\geq 0$ $\forall$ $l$, with all
$\Delta^-(l)$ of the {\it same} sign, the
strongest condition is obtained for $k=0$, so that stability occurs for
\begin{equation}\lambda>\lambda_c=\Delta^+_0+|\Delta^-_0|=
 \sum_l \Delta^+(l)+|\Delta^-(l)|\label{lc}\,.\end{equation}
For $\lambda\rightarrow\lambda_c$, $\omega_0\rightarrow 0$ (while all other
$\omega_k$ remain finite in a finite system), implying a divergence of
$f_0^\pm$ (Eq.\ (\ref{Fpm})):
\begin{equation}
|f^{-}_0|\approx \sqrt{\frac{|\Delta^-_0|}{8(\lambda-\lambda_c)}},
\;\;f^+_0\approx |f^-_0|-1/2\,,
\end{equation}
plus terms $O(\lambda/\lambda_c-1)^{1/2}$. This entails in turn a divergence
$f_0^A\propto (\lambda/\lambda_c-1)^{-1/4}$ of the largest eigenvalue
of a subsystem contraction matrix ${\cal D}_A$, with $S(\rho_A)\approx \ln
f^A_0+1\approx-\frac{1}{4}\ln(\lambda/\lambda_c-1)$ plus constant terms.

For example, the single site entropy (\ref{Si}) becomes
\begin{eqnarray}
S(\rho_i)&=&h(f),\;\;f=\sqrt{(\half+F_0^+)^2-(F_0^-)^2}-\half\,,\label{f}
\end{eqnarray}
with $F_0^{\pm}=\frac{1}{n}\sum_{k} f_k^\pm$ (Eq.\ (\ref{Fk})). For weak
coupling,
\begin{equation}
f\approx F_0^+-(F_0^-)^2\approx\frac{\sum_{l\neq 0}(\Delta^-(l))^2}
{4\lambda^2}\,,\label{fw}
\end{equation}
which involves just the couplings $\Delta^-(l)$ connecting the site with the
rest of the system. On the other hand, for  $\lambda\rightarrow
\lambda_c$, $f\propto\sqrt{\frac{f_0^-}{n}}\propto
(\lambda/\lambda_c-1)^{-1/4}$.

\subsection{Even-odd entanglement entropy}

\begin{figure}[t]
\centerline{\hspace*{0.25cm}\scalebox{.6}
{\includegraphics{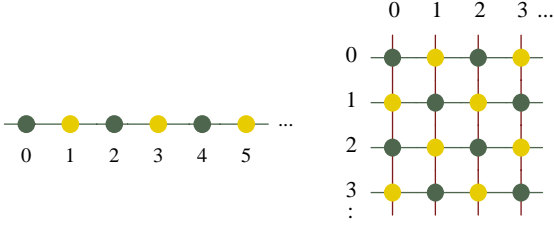}}}\vspace*{-.25cm} \caption{(Color online)
Even-odd partitions of one- and two-dimensional arrays} \label{f1}
\end{figure}

We now evaluate the entropy of the reduced state of all even sites,
$S(\rho_E)=S(\rho_{O})$, which measures their entanglement with the
complementary set of odd sites (Fig.\ \ref{f1} left). We will assume $n$ even,
such that the even subsystem, defined by $(-1)^i=+1$, is again translationally
invariant. The ensuing contraction matrix ${\cal D}_E$ can be obtained by
removing contractions between even and odd sites in the full matrix (\ref{D})
and extracting then the even part. This leads to elements
\begin{equation}
{\tilde{F}}^\pm_{ij}=\half F^\pm_{ij}(1+e^{i\pi (i-j)})\,,
\end{equation}
whose Fourier transforms are, using Eq.\ (\ref{Fk}),
\begin{equation}
\tilde{f}^{\pm}_k=\half(f^{\pm}_k+f^{\pm}_{k+n/2})\,.\label{ft}
\end{equation}
The final symplectic eigenvalues of ${\cal D}_E$ then become
\begin{eqnarray}
\tilde{f}_k&=&\sqrt{(\half+\tilde{f}^+_k)^2-(\tilde{f}^-_k)^2}]
-\half\,,\label{fE}
\end{eqnarray}
for $k=0,\ldots,n/2-1$. We then obtain
\begin{equation}
S(\rho_E)=\sum_{k=0}^{n/2-1}h(\tilde{f}_k)
=\half\sum_{k=0}^{n-1}h(\tilde{f}_k)\,.
 \label{SE}\end{equation}
Whenever $\tilde{f}_k$ can be approximated by a smooth function
$\tilde{f}(\tilde{k})$ of $\tilde{k}\equiv k/n$, we may replace (\ref{SE}) by
the integral
\begin{equation}
S(\rho_E)\approx \frac{n}{2}\int_{0}^{1}
h[\tilde{f}(\tilde{k})]d\tilde{k}\,.
 \label{SEc}\end{equation}
In these cases, we may then expect $S(\rho_E)$ {\it extensive}, i.e.,
proportional to the number $n/2$ of even sites. Let us remark, however, that
this is not always the case: In a completely and uniformly connected system
like the Lipkin model \cite{RS.80,BDV.06,WVB.10}, the contraction matrix will
have a {\it single} non-zero symplectic eigenvalue $f_{n_A}$ for {\it any}
subsystem \cite{MRC.10}, including the whole even set, and
$S(\rho_E)=h(f_{n/2})$ is no longer proportional to $n$. A similar lack of
extensivity holds in a finite system in the vicinity of the instability
($\lambda\rightarrow\lambda_c$, see below).

For {\it weak coupling}, Eqs.\ (\ref{wekk}), (\ref{wek}) and (\ref{ft}) lead to
\begin{eqnarray}
\tilde{f}_k&\approx &\frac{(\Delta_k^--\Delta_{k+n/2}^-)^2}{16\lambda^2}
=\frac{(\sum_{l\;{\rm odd}}e^{i2\pi kl/n}\Delta^-(l))^2}{4\lambda^2}\,,
 \label{fEw}\end{eqnarray}
which involves again just the couplings $\Delta^-(l)$ connecting the even and
odd subsystems. On the other hand, for $\lambda\rightarrow\lambda_c$
(Eq.\ (\ref{lc})), $\tilde{f}_0\approx
\half\sqrt{(1+2f_{n/2}^+)|f_0^-|-2f_{n/2}^-f_0^-}$ diverges as
$(\lambda/\lambda_c-1)^{-1/4}$ whereas all other $\tilde{f}_k$ remain finite,
and extensivity is lost.

\subsection{First neighbor coupling}
Let us now examine in detail the first neighbor case $\Delta^{\pm}(l)=\half
\Delta^{\pm}(\delta_{l1}+\delta_{l,-1})$, where  Eq.\ (\ref{Dk}) becomes
\begin{equation}
\Delta^{\pm}_k=\Delta^{\pm}\cos(2\pi k/n)\label{DC}\,.
 \end{equation}
The exact $S(\rho_E)$ can be obtained from Eqs.\ (\ref{ft})--(\ref{SE}).
In the weak coupling limit, Eqs.\ (\ref{fw}) and (\ref{fEw}) lead to
\begin{eqnarray}
f&\approx&\frac{(\Delta^-)^2}{8\lambda^2}\,,\label{fa}\\
\tilde{f}_k&\approx&\frac{(\Delta^-_k)^2}{4\lambda^2}\approx 2f\cos^2(2\pi
k/n)\,. \label{fka}
 \end{eqnarray}
Using Eqs.\ (\ref{Hw})--(\ref{SEc}), the single site and the total even
entropies can then be expressed just in terms of $f$:
 \begin{eqnarray}
S(\rho_i)&\approx & -f(\ln f -1)\,,\label{S0a}\\
S(\rho_E)&\approx&-nf\!\int_{0}^1\!\!\!\!\cos^2 (2\pi\tilde{k})
\{\ln[2f\cos^2(2\pi\tilde{k})]-1\}\,d\tilde{k}\nonumber\\
&=&-\frac{n}{2}f(\ln f-\ln 2)\,.\label{SEa}
 \end{eqnarray}
Hence, in this limit $S(\rho_E)$ is {\it extensive}, becoming $n/2$ times the
single site entropy (\ref{S0a}) minus a $O(nf)$ correction accounting for the
interaction between even sites:
\begin{equation} S(\rho_E)\approx \frac{n}{2}S(\rho_i)-\frac{n}{2}f(1-\ln 2)
 \,.\label{SEi}\end{equation}
The last term represents the {\it even mutual entropy}
$\frac{n}{2}S(\rho_i)-S(\rho_E)$, which is always a positive quantity and
becomes here also {\it extensive} in this limit.

In contrast, the block entropy $S(\rho_L)$, where $\rho_L$ denotes a contiguous
block of $L<n$ spins, rapidly saturates as $L$ increases \cite{AEPW.02}. In the
weak coupling limit, it is verified that the ensuing contraction matrix ${\cal
D}_L$ possesses, up to lowest non-zero order, just {\it two} positive non-zero
symplectic eigenvalues $f_L^{\pm}\approx\half f$ for {\it any} $L\geq 2$,  such
that
\begin{eqnarray}
S(\rho_L)&\approx& -f(\ln f/2-1)\label{SLw}\approx  S(\rho_i)+f\ln 2\,
 \label{SLw2}\,,\end{eqnarray}
for $2\leq L\leq n-2$, i.e., it saturates already for $L=2$.
Hence, in this limit,
\begin{equation}
 S(\rho_E)\approx \frac{n}{2}S(\rho_L)-\frac{n}{2}f\,.\label{SEL}
 \end{equation}

\begin{figure*}[t]
\centerline{\scalebox{.9}{\includegraphics{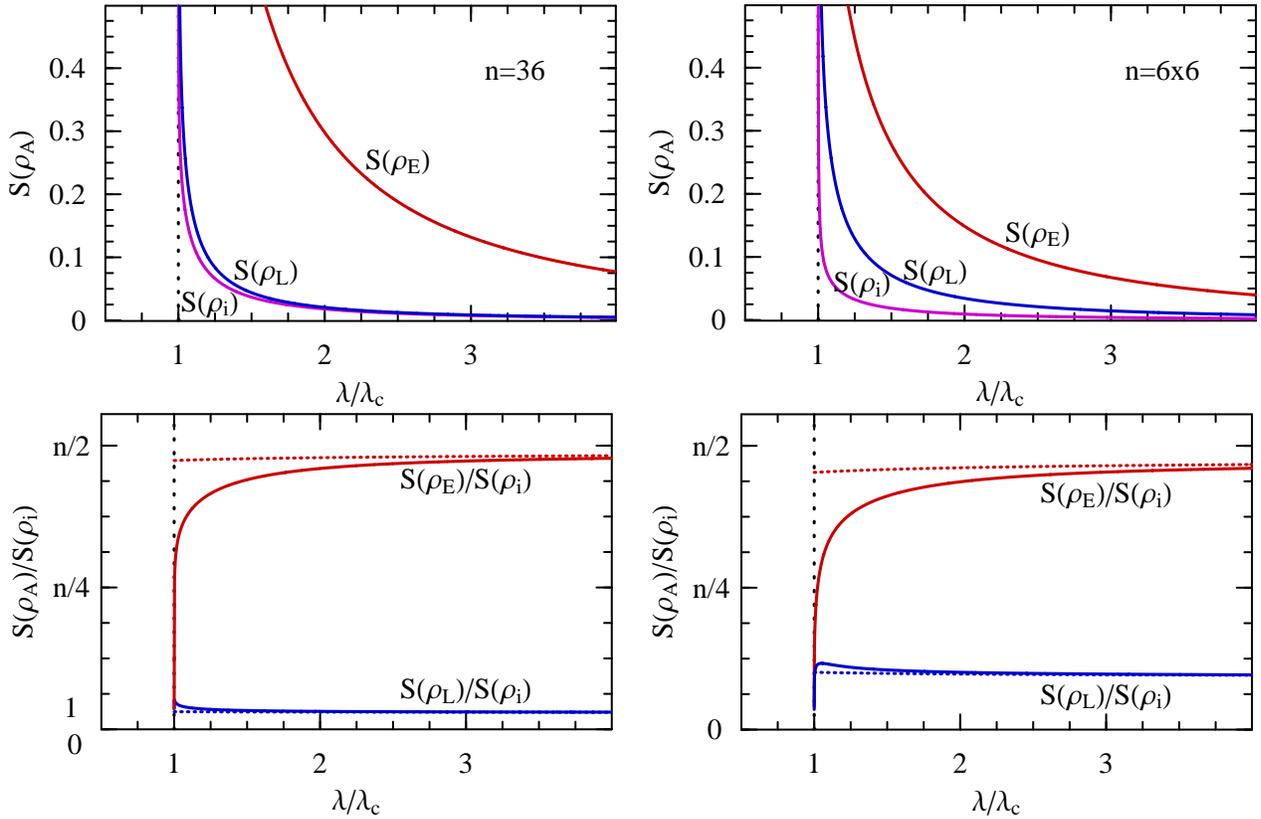}}}
\vspace*{-.25cm} \caption{(Color online) Top left: Entanglement entropies in
the ground state of a one-dimensional bosonic chain of $n=36$ sites described by
the Hamiltonian (\ref{H1}) with first neighbor cyclic couplings and
$\Delta^-=\Delta^+/3$. $S(\rho_i)$, $S(\rho_L)$ and $S(\rho_E)$ denote,
respectively, the entropy of the reduced state of a single site, a block of
$L=n/2$ contiguous sites and the set all even sites. Bottom left: The ratios
$S(\rho_E)/S(\rho_i)$ and $S(\rho_L)/S(\rho_i)$. Dotted lines depict the ratios
determined by the asymptotic expressions (\ref{S0a})--(\ref{SLw}), For large
$\lambda$ these ratios are then close to $n/2$ and $1$ respectively, while for
$\lambda\rightarrow\lambda_c$ they all approach 1. The right panels depict the
same quantities for a two-dimensional square array of $n=6\times 6$ sites with
isotropic cyclic couplings and the same ratio $\Delta^-/\Delta^+$. $S(\rho_L)$
denotes the entropy of a contiguous half of $6\times 3$ sites. Dotted lines
correspond to the expressions (\ref{SEb})--(\ref{SLb}). At fixed
$\lambda/\lambda_c$, $S(\rho_E)$ and $S(\rho_i)$ are now roughly half the value
of the left panel (Eq.\  (\ref{fd2})), while $S(\rho_L)/S(\rho_i)$  is
proportional to $\sqrt{n}$ (Eq.\ (\ref{SLb})).}
 \label{f2}
\end{figure*}

Assuming $\Delta^+>0$ (if $\Delta^+<0$ we can change its sign by a local change
$b_i\rightarrow -b_i$ at odd sites) the present system is stable for
$\lambda>\lambda_c=\Delta_++|\Delta_-|$ (Eq.\ (\ref{lc})). For
$\lambda\rightarrow\lambda_c$, $\omega_0\rightarrow 0$ and all
previous entropies diverge. In particular, Eq.\ (\ref{fE}) leads to
\[\tilde{f}_0\approx \half[\sqrt[4]{\frac{|\Delta^-|\lambda_c}
 {2^3\Delta^+(\lambda-\lambda_c)}}-1]\,,\]
being then verified that
$S(\rho_E)\approx -\frac{1}{4}\ln(\lambda/\lambda_c-1)$ plus a constant term up
to leading order. Hence, in this limit $S(\rho_E)/S(\rho_i)\rightarrow 1$.

As illustration, the left panels in Fig.\ \ref{f2} depict the single site,
block and even-odd entanglement entropies for a ring of $n=36$ sites with
$\Delta^-=\Delta^+/3$, where $\lambda_c=4\Delta^+/3$.

\subsection{Even-Odd entropy in d-dimensions}

The whole previous treatment can be directly extended to a translationally
invariant cyclic array in $d$ dimensions. We should just replace $l,k,n$ by
vectors $\bm{l}=(l_1,\ldots,l_d)$, $\bm{k}=(k_1,\ldots,k_d)$ and
$\bm{n}=(n_1,\ldots,n_d)$, with $l_i,k_i=0,\ldots, n_i-1$. We will assume
couplings satisfying $\Delta^{\pm}_{\bm{i},\bm{j}}=
\Delta^{\pm}(\bm{i}-\bm{j})$, with $\Delta^{\pm}(-\bm{l})=
\Delta^{\pm}(\bm{n}-\bm{l})=\Delta^{\pm}(\bm{l})$. The same previous
expressions (\ref{wk})--(\ref{Fpm}) then hold, with
\begin{eqnarray}
\Delta^{\pm}_{\bm k}&=&\sum_{\bm{l}}e^{i2\pi\tilde{\bm{k}}\cdot\bm{l}}
 \Delta^{\pm}(\bm{l})\,,\label{dkv}\\
F^{\pm}_{\bm{l}}&=&\frac{1}{n} \sum_{\bm{k}}e^{-i2\pi\tilde{\bm{k}}\cdot\bm{l}}
f^{\pm}_{\bm{k}},\label{fkv}
\end{eqnarray}
where $\tilde{\bm{k}}=(k_1/n_1,\ldots,k_d/n_d)$ and $n=\prod_{i=1}^d n_i$ is
the total number of sites. Eqs.\ (\ref{wekk})--(\ref{wek}) remain  unchanged
with $i,k,l\rightarrow \bm{i},\bm{k},\bm{l}$.

The subsystem of all even sites, like that formed by the blue sites in Fig.\
\ref{f1} right, is defined by
\[(-1)^{i_1+\ldots+i_d}=+1\,.\]
Its contraction matrix will then be the even block of
\begin{equation}
\tilde{F}^{\pm}_{\bm{ij}}=\half F^\pm_{\bm{i}\bm{j}}(1+e^{i\pi(\bm{i}-\bm{j})
 \cdot\bm{1}})\end{equation}
where $\bm{1}=(1,\ldots,1)$. Assuming $n_i$ even $\forall$ $i$, its Fourier
transform is then given again by
\begin{equation}
\tilde{f}^\pm_{\bm{k}}=\half[f^{\pm}_{\bm{k}}+f^{\pm}_{\bm{k}+\bm{n}/2}]\,,
\end{equation}
where $k_i+n_i/2\rightarrow k_i-n_i/2$ if $k_i\geq n_i/2$. The symplectic
eigenvalues of ${\cal D}_E$ are then given again by Eq.\ (\ref{fE}) with
$k\rightarrow \bm{k}$, and the even-odd entanglement entropy reads
\begin{eqnarray}
S(\rho_E)&=&\half \sum_{\bm{k}} h(\tilde{f}_{\bm{k}})\approx \frac{n}{2}\int
h[\tilde{f}(\bm{\tilde{k}})]d^d\tilde{k}\,, \label{SEd}
\end{eqnarray}
where $k_i=0,\ldots,n_i-1$ in the sum and the integral is restricted to the
unit cube $0\leq \tilde{k}_i\leq 1$ and valid if $\tilde{f}_{\bm{k}}$ is a
smooth function $\tilde{f}(\tilde{\bm{k}})$ of $\bm{\tilde{k}}$.

In the case of first neighbor couplings
\[\Delta^\pm(\bm{l})=\frac{1}{2}\sum_{i=1}^d
 \Delta^{\pm}_i(\delta_{\bm{l},\bm{e}_i}+\delta_{\bm{l},-\bm{e}_i})\,,\]
where $\bm{e}_i=(0,\ldots 1_i,\ldots 0)$, Eq.\ (\ref{dkv}) leads to
\begin{equation}
\Delta^{\pm}_{\bm{k}}=\sum_{i=1}^d \Delta^{\pm}_i\cos(2\pi k_i/n_i)\,.
\end{equation}
with $\Delta^\pm_{\bm{k}+\bm{n}/2}=-\Delta^\pm_{\bm{k}}$.
In the weak coupling limit we then obtain
\begin{eqnarray}
f&\approx &\frac{|\bm{\Delta}^-|^2}{8\lambda^2},\;\;
|\bm{\Delta}^-|^2=\sum_{i=1}^d (\Delta^{-}_i)^2\,,\label{fd}\\
\tilde{f}_{\bm{k}}&\approx &  u(\bm{\tilde{k}})\,f\,,
\;\;u(\bm{\tilde{k}})=2(\sum_i
\frac{\Delta^-_i}{|\bm{\Delta}^-|}\cos\frac{2\pi k_i}{n_i})^2\,.
\end{eqnarray}
Hence, the single site entropy is again $S(\rho_{\bm{i}})\approx -f(\ln f-1)$
while Eq.\ (\ref{SEd}) yields
\begin{eqnarray}
S(\rho_E)&\approx& -\frac{n}{2}f(\ln f-1+\alpha)\label{SEb}\\
&\approx&\frac{n}{2}S(\rho_{\bm{i}})-\frac{n}{2}\,f\,\alpha\,,\label{SEbi}
 \end{eqnarray}
where $\alpha$ is a {\it geometric} entropy factor:
\begin{equation}
\alpha=\int u(\bm{\tilde{k}})\ln u(\bm{\tilde{k}})\, d^d\tilde{k}\,,
\end{equation}
($u(\bm{\tilde{k}})\geq 0$, $\int u(\bm{\tilde{k}})d^d\tilde{k}=1$). In the
isotropic case $\Delta^-_i=\Delta^-$ $\forall$ $i$, we have $\alpha=\alpha_d$,
with $\alpha_1=1-\ln 2\approx 0.307$ (Eq.\ \ref{SEi}),
$\alpha_2=2\alpha_1\approx 0.614$ and $\alpha_3\approx 0.636$, approaching
$\approx \ln 2$ for large $d$.

At fixed $\lambda$, and for $\Delta^\pm_i=\Delta^\pm$,
$f=(\Delta^-)^2 d/(8\lambda^2)$ and hence  both $S(\rho_{\bm{i}})$ and
$S(\rho_E)$ increase as $d$ increases, reflecting the larger number of links.
However, and assuming again $\Delta^+\geq 0$, $\lambda_c=d(\Delta^++|\Delta^-|)$
also increases, entailing that at fixed $\lambda/\lambda_c$, $f$ (and so
$S(\rho_{\bm{i}})$ and $S(\rho_E)$) decreases:
\begin{equation}
f\approx\frac{[\Delta^-/(\Delta^++|\Delta^-|)]^2}{8d(\lambda/\lambda_c)^2}\;\;
 \,.\label{fd2}
 \end{equation}
For example, the right panels in Fig.\ \ref{f2} depict $S(\rho_E)$ and
$S(\rho_{\bm{i}})$ in an isotropic square lattice of $6\times 6$ sites, with
the same previous ratio $\Delta^-/\Delta^+=1/3$. At fixed $\lambda/\lambda_c$,
their values are verified to be roughly half that of the similar
one-dimensional case (Eq.\ (\ref{fd2})). Their ratio is also slightly smaller
due to the increase in the parameter $\alpha$ in  (\ref{SEbi}). On the other
hand, for $\lambda\rightarrow \lambda_c$ there is again a single vanishing
energy $\omega_{\bm{0}}$, so that all entropies behave as $-\frac{1}{4}\ln
(\lambda/\lambda_c-1)$ up to leading order, with all ratios approaching $1$.

We also depict there the entropy $S(\rho_L)$ of a contiguous  half-size block
($n_x\times n_y/2=6\times 3$ sites), which is now proportional to its boundary
$2n_x$. For $\lambda\gg \lambda_c$, it is verified that the number of non-zero
positive eigenvalues of the corresponding contraction matrix ${\cal D}_L$ is
just the number of couplings ``broken'' by the partition ($2 n_x$), being all
approximately equal to $f/4$ up to leading non-zero order. We then obtain
\begin{equation}
S(\rho_L)\approx -\frac{n_x}{2}f(\ln f/4-1)\approx \frac{n_x}{2}
[S(\rho_{\bm{i}})+2f\ln 2]\,,\label{SLb}
 \end{equation}
whence $S(\rho_L)/S(\rho_i)\propto n_x/2$ in this limit, as verified in the
right panels of Fig.\ \ref{f2}.

\section{Application to spin systems\label{III}}

The previous bosonic formalism can be directly applied to interacting spin $s$
systems in an external magnetic field through the RPA approximation
\cite{MRC.10}. Denoting with $s_{i\mu}$ the dimensionless spins
$S_{i\mu}/\hbar$ at site $i$, we will consider a cyclic translationally
invariant finite array which can be described by an $XY$ Hamiltonian of the
form
\begin{subequations}
\label{HS}
\begin{eqnarray}
H&=&B\sum_i s_{iz}-\frac{1}{2s}\sum_{i\neq j}(J^x_{ij}s_{ix}s_{jx}+
J^y_{ij}s_{iy}s_{jy})\label{Hs1}\\
&=&B\sum_i s_{iz}-\frac{1}{2s}[\sum_{i\neq j}\Delta^+_{ij}s_{i+}s_{j-}
+\half\Delta^-_{ij}(s_{i+}s_{j+}+h.c.)]\label{Hs2}
\end{eqnarray}\end{subequations}
where $s_{j\pm}=s_{jx}\pm i s_{jy}$, $J_\mu^{ij}=J_\mu(i-j)$ and
\begin{equation}
\Delta^{\pm}_{ij}=\half(J^x_{ij}\pm J^y_{ij})=\Delta^\pm (i-j)\,.\label{DS}
\end{equation}
We note that $x,y,z$ may in principle also denote local intrinsic axes at each
site, in which  case the field is assumed to be directed  along the local $z$
axis. The $s^{-1}$ scaling of the couplings ensures a spin-independent mean
field and effective RPA boson Hamiltonian (see below).

\begin{figure*}[!t]
\centerline{\scalebox{.9}{\includegraphics{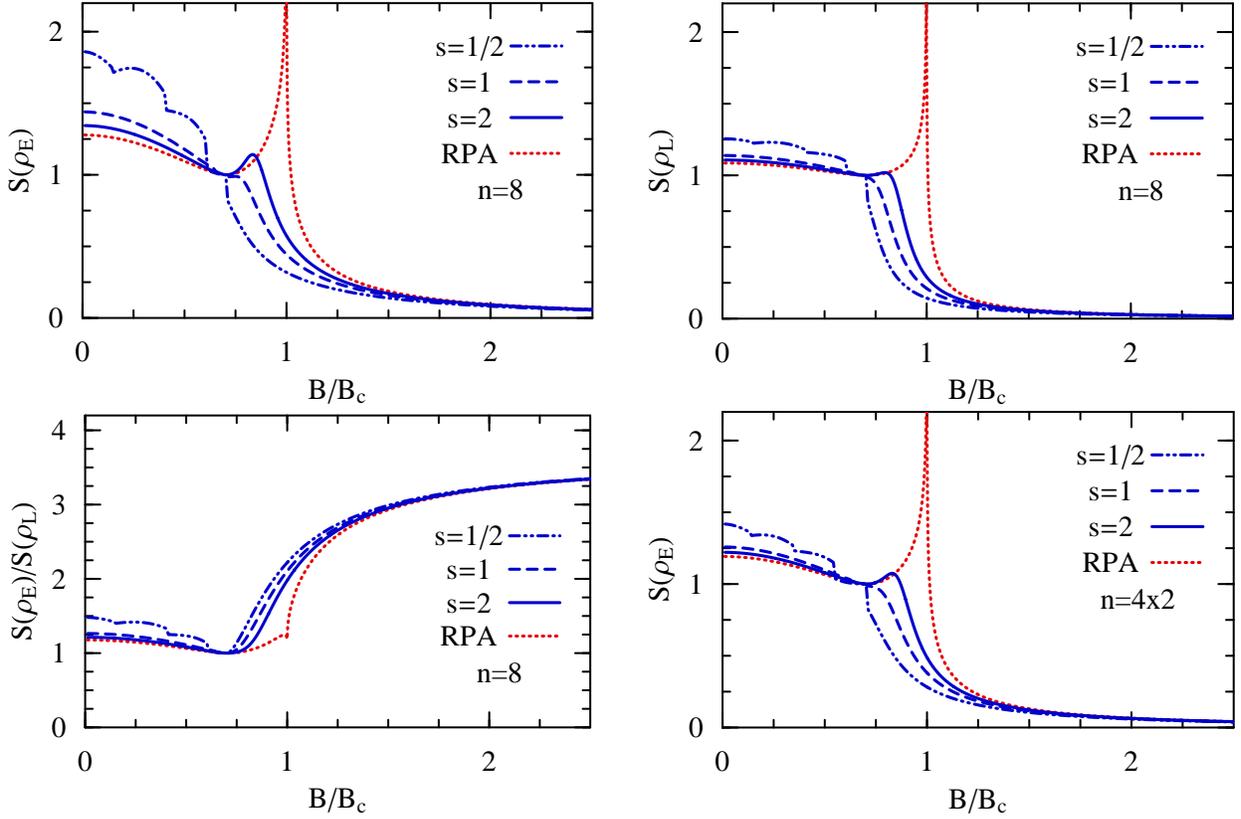}}}
\caption{(Color online) Top: Exact entanglement entropy of all
even sites (left) and of a contiguous block of $n/2$ sites (right) in the
ground state of a one dimensional cyclic chain of $n=8$ spins with anisotropic
$XY$ first neighbor couplings ($J_y/J_x=\half$) and spin $s=1/2$, $1$ and $2$,
as a function of the transverse magnetic field. The dotted line depicts the
bosonic RPA result, with $B_c=J_x$ the mean field critical field. We have used
base 2 logarithm in the entropy, such that all entropies approach $1$ at the
factorizing field $B_s\approx 0.71 B_c$. Bottom: Left: The corresponding ratio
$S(\rho_E)/S(\rho_L)$. Right: The entanglement entropy of all even sites in a
rectangular lattice of $4\times 2$ spins. Remaining details as in the top
panels.}
 \label{f3}
\end{figure*}

{\it Normal RPA}. For sufficiently strong field $B$, the lowest mean field
state (i.e., the separable state with lowest energy) is the aligned state
$|0\rangle=|0_1\rangle\otimes \ldots\otimes|0_n\rangle$, where $|0_i\rangle$
denotes the local state with maximum spin  along the $-z$ axis
($s_{iz}|0_i\rangle=-s|0_i\rangle$). In such a case, RPA implies the
approximate bosonization \cite{MRC.10}
\begin{equation}
s_{i+}\rightarrow \sqrt{2s}b^\dagger_i, \;\;s_{i-}\rightarrow
\sqrt{2s}b_{i}\,,\;\;s_{iz}\rightarrow b^\dagger_{i}b_i-\half\,,
 \label{bos}\end{equation}
which is similar to the Holstein-Primakoff bosoniztion \cite{RS.80,BDV.06} and
leads to the quadratic boson Hamiltonian (\ref{H1}) with the parameters
(\ref{DS}) and $\lambda=B$. We may then directly apply all previous expressions.

The  bosonic RPA scheme becomes exact for strong fields $|B|\gg B_c$ for any
size $n$, spin $s$, geometry or interaction range, since for weak coupling it
corresponds to the exact first order perturbative expansion of the ground state
wave function \cite{MRC.10}. As a check, in the case of the spin $1/2$
one-dimensional chain with first neighbor $XY$ coupling, an analytic expression
of the block entropy in the limit $n\rightarrow \infty$ has been obtained in
\cite{FIJK.07,IJK.05,PE.04}. For $\lambda=B>\Delta^+$, it is given in present
notation by \cite{FIJK.07}
\begin{equation}
S(\rho_L)={\textstyle\frac{1}{6}[\ln
\frac{4}{\alpha\alpha'}+(\alpha^2-\alpha'^2)\frac{2I(\alpha)I(\alpha')}{\pi}]},
\label{Sx}\end{equation} where
$\alpha=\Delta^-/\sqrt{\lambda^2+{\Delta^-}^2-{\Delta^+}^2}$,
$\alpha'=\sqrt{1-\alpha^2}$ and $I(\alpha)=\int_0^1 dx/\sqrt{(1-x^2)(1-\alpha^2
x^2)}$ is the elliptic integral of the first kind. An expansion of
(\ref{Sx}) for $\lambda\gg \Delta^\pm$ leads exactly to present Eq.\ (\ref{SLw}),
with $f$ given by (\ref{fa}). We can then expect the asymptotic expressions
(\ref{SEa}) and (\ref{SEbi}) for $S(\rho_E)$ to be exact in this limit also in
spin systems.

{\it Parity breaking RPA}. Considering now the anisotropic ferromagnetic-type
case $|J_y(l)|\leq J_x(l)$ $\forall$ $l$ in (\ref{Hs1}), the previous normal
RPA scheme will hold, according to Eq.\ (\ref{lc}), for $B\geq B_c=J_x^0\equiv
\sum_l J_x(l)$, i.e., when the corresponding boson system is stable.

For $|B|<B_c$, the normal RPA becomes unstable ($\omega_0$ becomes imaginary).
The lowest mean field state corresponds here to degenerate states
$|\pm\Theta\rangle$ fully aligned along an axis $z'$ forming an angle
$\pm\theta$ with the $z$ axis in the $x,z$ plane:
$|\Theta\rangle=|\theta_1\rangle\otimes\ldots\otimes|\theta_n\rangle$, with
$|\theta_i\rangle=\exp[-i\theta s_{iy}]|0_i\rangle$. We are assuming here an
anisotropic $XY$ coupling such that $H$ commutes with the $S_z$ parity
$P_z=e^{i\pi (\sum_i s_{iz}+ns)}$, but not with an arbitrary rotation around
the $z$ axis (as in the $XX$ case). Such states break then parity symmetry,
satisfying $P_z|\Theta\rangle=|-\Theta\rangle$. The angle $\theta$ is to be
determined from \cite{MRC.10}
\begin{equation}\cos\theta=B/B_c,\;\;B_c=\sum_l J_x(l)\,.\end{equation}
For $|B|<B_c$, the bosonization (\ref{bos}) is then to be applied in the RPA to
the rotated spin operators $s_{iz'}=s_{iz}\cos\theta+s_{ix}\sin\theta$,
$s_{i\pm'}=s_{ix'}\pm is_{iy'}$, with
$s_{ix'}=s_{ix}\cos\theta-s_{iz}\sin\theta$ and $s_{iy'}=s_{iy}$. This leads
again to a {\it stable} Hamiltonian of the form (\ref{H1}) with \cite{MRC.10}
\begin{equation}
\lambda=B_c\,,\;\;
\Delta^{\pm}(l)=\half[J^x(l)\cos^2\theta\pm J^y(l)]\,.\label{Dl}
 \end{equation}

For $|B|<B_c$ we should also take into account the important effects from
parity restoration for a proper RPA estimation of entanglement entropies
\cite{MRC.10}.  The exact ground state in a finite array will have a definite
parity $P_z$ outside crossing points \cite{RCM.08}, implying that the actual
RPA ground state should be taken as a definite parity superposition of the RPA
spin states constructed around $|\pm\Theta\rangle$ \cite{MRC.10}. This leads to
reduced RPA spin densities of the form
$\rho_A\approx\half[\rho_A(\theta)+\rho_A(-\theta)]$ if the complementary
overlap $O_{\bar{A}}=\langle -\Theta_{\bar{A}}|\Theta_{\bar{A}}\rangle$ can be
neglected. If the subsystem overlap
$O_A=\langle-\Theta_A|\Theta_A\rangle=\cos^{2n_A s}\theta$ is also negligible,
such that $\rho_A(\theta)\rho_A(-\theta)\approx 0$, then \cite{MRC.10}
 \begin{equation}S(\rho_A)\approx S(\rho_A(\theta))+\delta\,,
 \label{dl}\end{equation}
where $\delta=\ln 2$.  The final effect is then the addition of a constant
shift to the bosonic subsystem entropy for $|B|<B_c$. This is applicable to
both $S(\rho_E)$ and $S(\rho_L)$ if $\theta$, $n$ and the block
size $L$ are not too small.

For first neighbor couplings with anisotropy $\chi=J_y/J_x\in(0,1)$ (if
$\chi>1$ we just redefine the $x,y$ axes) as well as for arbitrary range
couplings with a common anisotropy $\chi=J_y(l)/J_x(l)\in(0,1)$, another
fundamental feature for $|B|<B_c$ is the existence of a transverse factorizing
field $B_s=B_c\sqrt{\chi}$ where the  mean field states $|\pm\Theta\rangle$
become exact ground states \cite{KTM.82,AA.06,RCM.08,GAI.08}. As seen from
(\ref{Dl}), at this field $\Delta^-(l)=0$ $\forall$ $l$, so that the RPA vacuum
remains the same as the mean field vacuum \cite{MRC.10} and all contractions
$F^\pm_{ij}$ vanish, implying $S(\rho_A(\theta))=0$. All RPA entropies at $B_s$
reduce then to the correction term $\delta$ arising from parity restoration
\cite{MRC.10}.

This is essentially also the exact result at $B_s$: The transverse factorizing
field corresponds to the last ground state parity transition as $B$ increases
from 0 \cite{RCM.08} and the ground state side-limits for $B\rightarrow B_s$
are actually the definite parity combinations of the mean field states
$|\pm\Theta\rangle$ \cite{RCM.08,RCM.09}. These definite parity states have
Schmidt number $2$ for any bipartition, implying that the side-limits of the
{\it exact} entropy of the reduced state of any subsystem at $B_s$ do not
approach 0 but rather the values \cite{RCM.09}
\begin{equation}S(\rho_A^\pm(B_s))=\sum_{\nu=\pm}q_\nu^\pm\ln q_\nu^\pm\,,\;
q_\nu^\pm=\frac{(1+\nu O_A)(1\pm\nu O_{\bar{A}})}
{2(1\pm O_A O_{\bar{A}})}\,,
 \label{Os}\end{equation}
where $+$ ($-$) corresponds to positive (negative) parity, i.e. the right
(left) side limit at $B_s$ \cite{RCM.09}. Eq.\ (\ref{Os}) is valid for any size
or spin. For small complementary overlap $\bar{O}_{\bar{A}}$, $q_\nu^\pm\approx
\half(1+\nu O_A)$ and both side limits coincide, while if $O_A$ is also small,
$q_\nu\approx 1/2$ and  Eq.\ (\ref{Os}) reduces to  $+\ln 2$. This is also in
agreement with the exact limit of the block entropy of the large one dimensional
$s=1/2$ $XY$ chain at $B_s$ \cite{FIJK.07}.
\begin{figure}[!t]
\centerline{\hspace*{0.25cm}\scalebox{.9}{\includegraphics{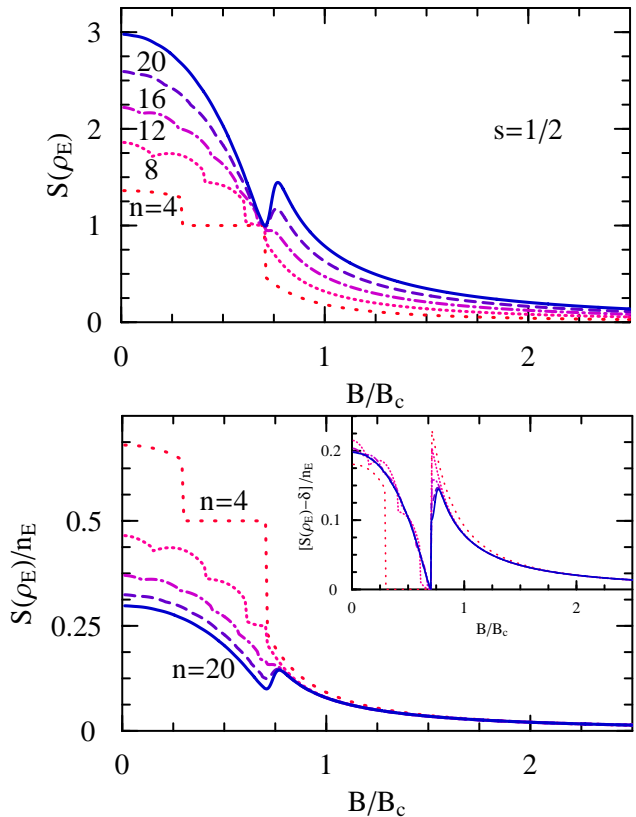}}}
\vspace*{-1.cm} \caption{(Color online) Top: Exact entanglement entropy of all
$n/2$ even sites in the ground state of a spin $1/2$ cyclic chain for different
values of $n$. Couplings are the same as in Fig.\ \ref{f3}. Bottom: The
intensive entropy $S(\rho_E)/n_E$ ($n_E=n/2$). All curves coalesce for
$B\agt\Delta_+ $. The inset depicts the intensive shifted entropy
$(S(\rho_E)-\delta)/n_E$, where $\delta=0$ for $B>B_s$ and $\delta=1$ for
$B<B_s$, which makes curves for $n\geq 8$ coalesce also for $B<B_s$.}
 \label{f4}
\end{figure}

Illustrative exact results for the even-odd entanglement entropy in a finite
linear cyclic spin $s$ chain with first neighbor couplings are plotted in the
top left panel of Fig.\ \ref{f3} for spins $s=1/2$, $1$ and $2$, together with
the bosonic RPA estimation. The exact definite parity ground state was employed
in all cases. We also depict for comparison the entropy of a contiguous half
(top right), and the ratio $S(\rho_E)/S(\rho_L)$ (bottom left). The anisotropy
of the coupling is the same as in Fig.\ \ref{f2} ($\Delta_-/\Delta_+=1/3$). The
RPA result (independent of $s$ for the scaling used in (\ref{HS})), represents
the large spin limit but is already quite close to the exact results for $s=2$
except in the vicinity of $B_c$, where the exact entropies remain of course
finite in a finite chain. The ratio $S(\rho_E)/S(\rho_L)$ is nonetheless quite
accurately reproduced and shows the extensive character of $S(\rho_E)$ for
$B>B_c$, in agreement with (\ref{SEL}), where the entropies for all spin values
rapidly approach the RPA result and become spin independent. For $|B|<B_c$ the
shift $\delta$ in (\ref{dl}) ($\delta=+1$ in Figs\ \ref{f3}--\ref{f4} since
base $2$ logarithm was employed) is essential for the agreement and explains
the lack of direct extensivity in this region. The collapse of all entropies to
the value $\delta$ at $B_s$ is also verified, and for $s=1/2$ even the small
discontinuity at $B_s$ predicted by Eq.\ (\ref{Os}) can be appreciated
(together with the other parity transitions for $B<B_s$). The bottom right
panel depicts $S(\rho_E)$ in a $4\times 2$ square lattice with identical
couplings in both directions and the same ratio $\Delta_-/\Delta_+$, where a
similar behavior is obtained. Exact results for $s=2$ are now even closer to
the RPA prediction, indicating that the accuracy of the latter tends to
improve, for stable mean fields, as the connectivity increases \cite{MRC.08}.

Exact results for a spin $1/2$ chain for different sizes are depicted in Fig.\
\ref{f4}. Even though RPA is not accurate for such low spin with a first
neighbor coupling \cite{MRC.08}, the exact results are again in qualitative
agreement with its predictions away from the critical region: Direct
extensivity $S(\rho_E)\propto n$ is verified for strong fields $B\agt\Delta_+$
(bottom panel), whereas for $B<B_s$ it holds for the shifted entanglement
entropy $S(\rho_E)-\delta$, as seen in the inset. Complete lack of extensivity
takes place at the factorizing field $B_s$, where the discontinuity implied by
(\ref{Os}) is appreciable for $n=8$ and becomes quite noticeable for $n=4$.

\section{Conclusions\label{IV}}

We have shown that the total even-odd entanglement entropy displays a strict
extensive behavior in both bosonic and spin chains or lattices for weak first
neighbor couplings (i.e., strong fields in a spin chain), providing explicit
asymptotic expressions for the general $d$ dimensional case. Extensivity of the
associated mutual information is also implied by these expressions. Deviations
from this behavior, however, were shown to arise for stronger couplings, i.e.,
proximity to the instability in the finite bosonic case or low fields in the
spin case. In the latter, a constant shift is essential to understand the exact
results for $|B|<B_c$, which has an evident meaning as a symmetry restoration
effect in the RPA. Besides, full loss of extensivity occurs in the
vicinity of the factorizing field. Present results confirm the validity of the
RPA approach (with inclusion of symmetry-restoration effects) for obtaining a
simple direct understanding of the main aspects of ground state entanglement in
spin chains, at least in those regions where a well defined mean field minimum
exists.

The authors acknowledge support from CIC (RR) and CONICET (NC,JMM) of Argentina.


\begin{thebibliography}{999}
\bibitem{NC.00}M.A.\ Nielsen and I.L.\ Chuang, {\it Quantum Computation and
             Quantum Information} (Cambridge Univ. Press, Cambridge, UK, 2000).
\bibitem{ON.02} T.J.~Osborne, M.A.~Nielsen, Phys.\  Rev.\ A {\bf  66},
                032110 (2002).
\bibitem{AFOV.08} L.~Amico, R.~Fazio, A.~Osterloh and V.~Vedral,
             Rev.~Mod.~Phys.\ {\bf 80}, 517 (2008).
\bibitem{ECP.10}J.\ Eisert, M.\ Cramer, M.B.\ Plenio,
Rev.\  Mod.\  Phys.\ {\bf 82}, 277 (2010).
\bibitem{BE.93} C.H.\ Bennett et al., Phys.\ Rev.\ Lett.\ {\bf 70}, 1895
    (1993); Phys.\ Rev.\ Lett.\ {\bf 76}, 722 (1996).
\bibitem{JLV.03} R.\ Josza and N.\ Linden, Proc.\ R.\ Soc.\
{\bf A 459}, 2011 (2003);
G.\ Vidal, Phys.\ Rev.\ Lett.\ {\bf 91}, 147902 (2003).
\bibitem{RB.01}R.\ Raussendorf and H.J.\ Briegel, Phys.\ Rev.\ Lett.\ 86,
5188 (2001); R.\ Raussendorf, D.E.\ Browne and H.J.\ Briegel,
Phys.\ Rev.\ A 68, 022312 (2003).
\bibitem{OAFF.02}A. Osterloh, L.\ Amico, G. Falci, R. Fazio,
Nature 416, 608 (2002).
\bibitem{VLRK.03} G.~Vidal, J.I.~Latorre, E.~Rico and A.~Kitaev,
 Phys.\ Rev.\ Lett.\ {\bf 90}, 227902 (2003).
 \bibitem{PEDC.05} M.B.\ Plenio, J.\ Eisert, J.\ Drei\ss ig, M.\ Cramer,
 Phys.\ Rev.\ Lett.\ {\bf 94}, 060503 (2005).
 \bibitem{AEPW.02} K.\ Audenaert, J.\ Eisert, M.B.\ Plenio, R.F.\ Werner,
 Phys.\ Rev.\ A {\bf 66} 042327 (2002).
\bibitem{CC.04}P.\ Calabrese and J.\ Cardy, J.\ Stat.\ Mech.\  P06002 (2004).
\bibitem{FIJK.07} F.\ Franchini, A.R.\ Its, B-Q.\ Jin, V.\ Korepin,
J.\ Phys.\ A {\bf 40} 8467 (2007).
\bibitem{IJK.05} A.R.\ Its, B-Q.\ Jin, V.\ Korepin,
J.Phys.\ A  {\bf 38} 2975 (2005).
\bibitem{PE.04} I.\ Peschel, J.\ Stat.\ Mech.\ P12005 (2004).
\bibitem{CWZ.06} Y.\ Chen, Z.D.\ Wang and F.C.\ Zhang,
Phys.\ Rev.\ B {\bf 73}, 224414 (2006).
\bibitem{KMN.06} J.P.\ Keating, F.\ Mezzadri, M.\ Novaes,
Phys.\ Rev.\ A {\bf 74} 012311 (2006).
\bibitem{MRC.10} J.M.\ Matera, R. Rossignoli, N.\ Canosa,
Phys.\ Rev.\ A {\bf 82}, 052332 (2010).
\bibitem{KTM.82}J.\ Kurmann, H.\ Thomas, G.\ M\"uller, Physica A
 {\bf 112}, 235 (1982).
\bibitem{AA.06} L.\ Amico  et al, Phys.\ Rev.\ A {\bf 74},
 022322 (2006);
 F.\ Baroni et al, J.\ Phys.\ A {\bf 40} 9845 (2007).
\bibitem{RCM.08} R.\ Rossignoli, N.\ Canosa, J.M. Matera,
  Phys.\ Rev.\ A  {\bf 77}, 052322 (2008).
\bibitem{GAI.08} S.M.\ Giampaolo, G.\ Adesso, F.\ Illuminati,
 Phys.\ Rev.\ Lett.\ {\bf 100}, 197201 (2008);
 Phys.\ Rev.\ B {\bf 79}, 224434 (2009).
\bibitem{RS.80} Peter Ring and Peter Schuck,
{\it The Nuclear Many-Body Problem} (Springer-Verlag, NY, 1980).
\bibitem{CEPD.06}M.~Cramer,  J.~Eisert, M.B.~Plenio, J.~Drei\ss ig,
 Phys.\ Rev.\ A {\bf 73} 012309 (2006).
\bibitem{ASI.04} G.~Adesso, A.~Serafini, F.~Illuminati,
Phys.~Rev.~A {\bf 70}, 022318 (2004); A.~Serafini, G.~Adesso,
F.~Illuminati, Phys.~Rev.~A {\bf 71}, 032349 (2005); G.~Adesso,
F.~Illuminati, Phys.~Rev.~A {\bf 78}, 042310 (2008).
\bibitem{BDV.06} T.\ Barthel, S.\ Dusuel, and J.\ Vidal,
Phys.\ Rev.\ Lett. {\bf 97}, 220402 (2006);
S.\ Dusuel and J.\ Vidal, Phys.\ Rev.\ B {\bf 71}, 224420 (2005);
J\ Vidal, S.\ Dusuel, and T.\ Barthel,
               J.\ Stat.\ Mech.\ P01015 (2007).
\bibitem{WVB.10} H.~Wichterich, J.~Vidal, and  S.~Bose
Phys.\ Rev.\ A {\bf 81}, 032311 (2010).
\bibitem{RCM.09} R.~Rossignoli, N.~Canosa, and J.M.~Matera,
Phys.\ Rev.\ A {\bf 80}, 062325 (2009).
\bibitem{MRC.08} J.M.\ Matera, R.~Rossignoli, N.~Canosa,
Phys.\ Rev.\ {\bf A 78}, 042319 (2008).
\end{thebibliography}
\end{document}